\documentclass{svproc}
\usepackage{url}

\usepackage{ amssymb }
\usepackage{amsmath}
\usepackage{slashed}
\usepackage{braket}
\usepackage{graphicx}
\usepackage{sidecap}
\usepackage{multicol}
\usepackage{wrapfig}
\graphicspath{ {images/} }
\usepackage{float}
\begin{document}
\mainmatter              
\title{A Beyond Mean Field Approach to Yang-Mills Thermodynamics}
\titlerunning{Beyond Mean Field Approach to Y-M Thermodynamics}  
%
\author{Pracheta Singha\inst{1,2} \and Rajarshi Ray\inst{1}\and
Chowdhury Aminul Islam \inst{3}\and Munshi G Mustafa\inst{4}}
\authorrunning{Pracheta Singha et al.} 
%
%
\institute{Center for Astroparticle Physics \& Space Science, Bose
Institute, Block-EN, Sector-V, Salt Lake, Kolkata-700091, India\\
\and
\email{pracheta.singha@jcbose.ac.in}
\and
School of Nuclear Science and Technology, University of Chinese Academy of Sciences, Beijing 100049, China
\and
Theory Division, Saha Institute of Nuclear Physics,
1/AF, Bidhannagar, Kolkata 700064, India
}
\maketitle              

\begin{abstract}
We propose a beyond mean field approach to evaluate Yang-Mills
thermodynamics from the partition function with
n-body gluon contribution, in the presence of a uniform background
Polyakov field. Using a path integral based formalism , we obtain,
unlike the previous mean field studies within this model framework, physically consistent 
results with good agreement to the lattice data throughout the
temperature range. 
\end{abstract}
\section{Introduction}
%
Thermodynamic quantities are some of the
fundamental observables to understand the properties of the strongly 
interacting medium and the nature of the associated phase transition. For
 SU(3) pure gauge theory the center symmetry, a global Z(3)    
symmetry, is spontaneously broken at 
high temperatures resulting in a first-order phase transition from the
 confined phase to the plasma phase~\cite{Karsch:1989pn}. The order parameter for this phase 
transition is the average of Polyakov loop, defined by,
\begin{equation}
L(\vec{x})=\frac{1}{3}\text{Tr}\bigg(\mathcal{T}\text{exp}[ ig \int_{0}^{\beta}{d\tau \mathcal{A}_0
(\vec{x},\tau)}]\bigg)
\label{polyakov}
\end{equation}
\noindent
where, $\mathcal{A}_0(\vec{x},\tau)$ is the temporal component of the 
gluonic field and $\tau$ is the Euclidean time.\\
To study SU(3) pure gauge system, we considered a partition function of thermal gluons in the presence of a background Polyakov field~\cite{Sasaki:2012bi,Islam:2012kv}, but instead of the mean field
 approximation, which may lead to the unphysical results~\cite{Sasaki:2012bi,Islam:2012kv}, we took a beyond mean field approach based on the path integral formalism to achieve a physically consistent model framework
with gluonic distribution function. In Sec.(\ref{sec:form}) we describe
our formalism followed by results and discussion in Sec.(\ref{sec:result}).

\section{\label{sec:form}Formalism}
The thermodynamic description of a gluonic quasiparticle system with a
  background Polyakov field can be formulated using the partition 
  function \cite{Sasaki:2012bi,Islam:2012kv},
\begin{eqnarray}
Z=\int\prod_{\bf x}  d\theta_3({\bf x}) d\theta_8({\bf x})
\text{Det}_{VdM} \text{exp}\left({-2V\int{\frac{d^3p}
{(2\pi)^3}\text{ln}\bigg{(}1+\sum^{8}_{n=1} a_n e^{-\frac{n |\vec{p}|}
{T}}\bigg{)}}}\right)~,
\label{partLA}
\end{eqnarray}

\noindent
where  $\theta_3$ and $\theta_8$, are two independent  parameters that
characterize the SU(3) group elements and can be associated with two diagonal generators $T_3$ and $T_8$.\\
 The coefficients $a_n$ for $n= 1 \cdots 8$, are the following,
\begin{eqnarray}
&a_1 = a_7=1-9\bar{\Phi}\Phi;\hspace{2mm}
a_2= a_6=1-27\bar{\Phi}\Phi;\hspace{2mm} 
a_3= a_5=-2+27\bar{\Phi}\Phi-81(\bar{\Phi}\Phi)^2\nonumber\\
& a_4= 2[-1+9\bar{\Phi}\Phi-27(\bar{\Phi}^3+\Phi^3)+81(\bar{\Phi}\Phi)^2];
\hspace{2mm} a_8= 1~.
\label{polya_distri}
\end{eqnarray}
Where, $\Phi$ and $\bar{\Phi}$ are the normalised characters defined as,
\begin{equation}
\Phi=\frac{1}{N_c}\text{Tr}\hat{L}_F; \hspace{5mm}    
\bar{\Phi}=\frac{1}{N_c}\text{Tr}\hat{L}^\dagger_F~,
\label{funcharacter}
\end{equation}
with, $\hat{L}_F$, the Polyakov line in the fundamental representation, 
given as,
\begin{equation}
\hat{L}_F = \text{diag}(e^{i\theta_3},e^{i\theta_8},e^{-i(\theta_3+\theta_8)})~.
\label{funmatrix}
\end{equation}
$\text{Det}_{VdM}$ is the Vandermonde determinant 
\cite{Sasaki:2012bi,Islam:2012kv}, given by, 
\begin{eqnarray}
\text{Det}_{VdM} &=& 64
\sin^2\frac{(\theta_3-\theta_8)}{2} \sin^2\frac{(2\theta_3+\theta_8)}{2}
\sin^2\frac{(\theta_3+2\theta_8)}{2} ~.
\label{haarphi}
\end{eqnarray}
Next, to obtain the thermodynamic observables, instead of evaluating the infinite dimensional integration in Eqn.(\ref{partLA}),  one may use 
the saddle point approximation. Here,  solving the following equations,
\begin{eqnarray}
\frac{\partial \Omega}{\partial \Phi} =0 ~;~ 
\frac{\partial \Omega}{\partial \bar{\Phi}}=0~,
\label{saddle}
\end{eqnarray}  
one can obtain the mean fields $\Phi_{mf}$ and $\bar{\Phi}_{mf}$ and all the thermodynamic observables in terms of those fields. 
 However, the thermodynamic quantities evaluated from this $\Omega=\Omega(\Phi_{mf},\bar{\Phi}_{mf})$  show unphysical
 behaviour below the transition temperature~\cite{Sasaki:2012bi,Islam:2012kv}, which we claim to be an artefact of
 using the mean field approximation. Noting that the thermodynamic 
 potential, in terms of the Polyakov loop,  is an oscillating function of
 $\theta_3$ and $\theta_8$, we propose a beyond mean field approach to include the significant contribution from the configurations away from the
 mean field. As the Polyakov loop fields in Eqn.(\ref{partLA}) are
  spatially
 uniform, we consider the configuration space to consist of $N
\rightarrow \infty$ points and  defining, 
\begin{eqnarray}
z =\int d\theta_3 d\theta_8
\text{Det}_{VdM} \text{exp}\left({-2\delta v \int{\frac{d^3p}{(2\pi)^3}
\text{ln}\big{(}1+\sum^{8}_{n=1} a_n e^{-\frac{n |\vec{p}|}
{T}}\big{)}}}\right)~,
\label{eq.z}
\end{eqnarray}
where $\delta v$ is a parameter with the dimension of volume, we write Eqn.(\ref{partLA}) as,
\begin{equation}
Z=z^N
\end{equation}
 The
thermodynamic variables then follow simply from the pressure which is
given as,
\begin{eqnarray}
p = \frac{T}{V}\text{ln}[Z] = \frac{T}{N\delta v} N\text{ln} z
= \frac{T}{\delta v} \text{ln} z~.
\label{eq.pressure1}
\end{eqnarray}
The expectation values of the local operators are obtained as,
\begin{eqnarray}
<O[\Phi({\bf x}),\bar\Phi({\bf x})]> = &&\frac{1}{z}\int  d\theta_3({\bf x}) d\theta_8({\bf x})
\text{Det}_{VdM} O[\Phi({\bf x}),\bar\Phi({\bf x})]
\nonumber \\
&& \text{exp}\left({-2\delta v\int{\frac{d^3p}
{(2\pi)^3}\text{ln}\big{(}1+\sum^{8}_{n=1} a_n e^{-\frac{n |\vec{p}|}
{T}}\big{)}}}\right)~.
\label{eq.expect2}
\end{eqnarray}
\section{\label{sec:result} Result and Discussion}
\begin{figure}[H]
\centering
  \includegraphics[height=1.5in,width =2.0in]{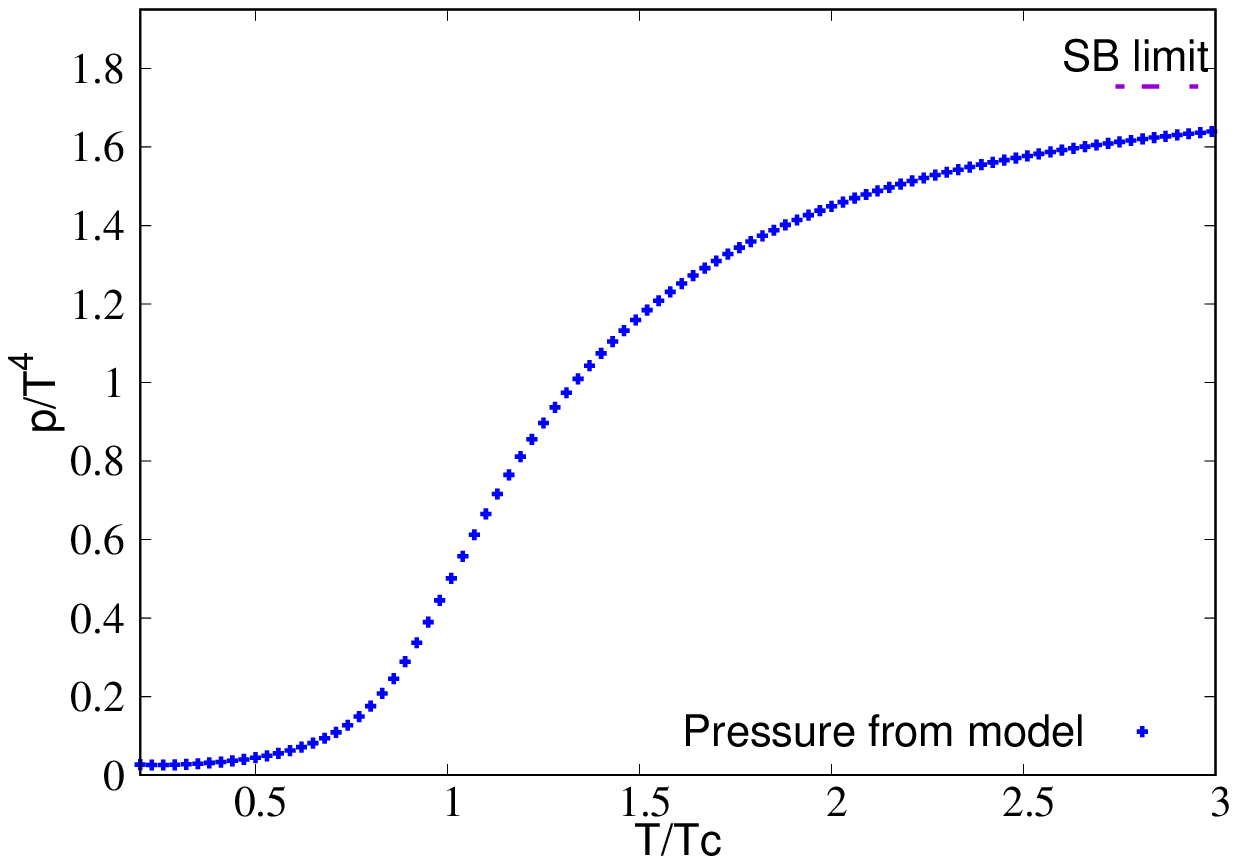}
  \includegraphics[height=1.5in,width =2.0in]{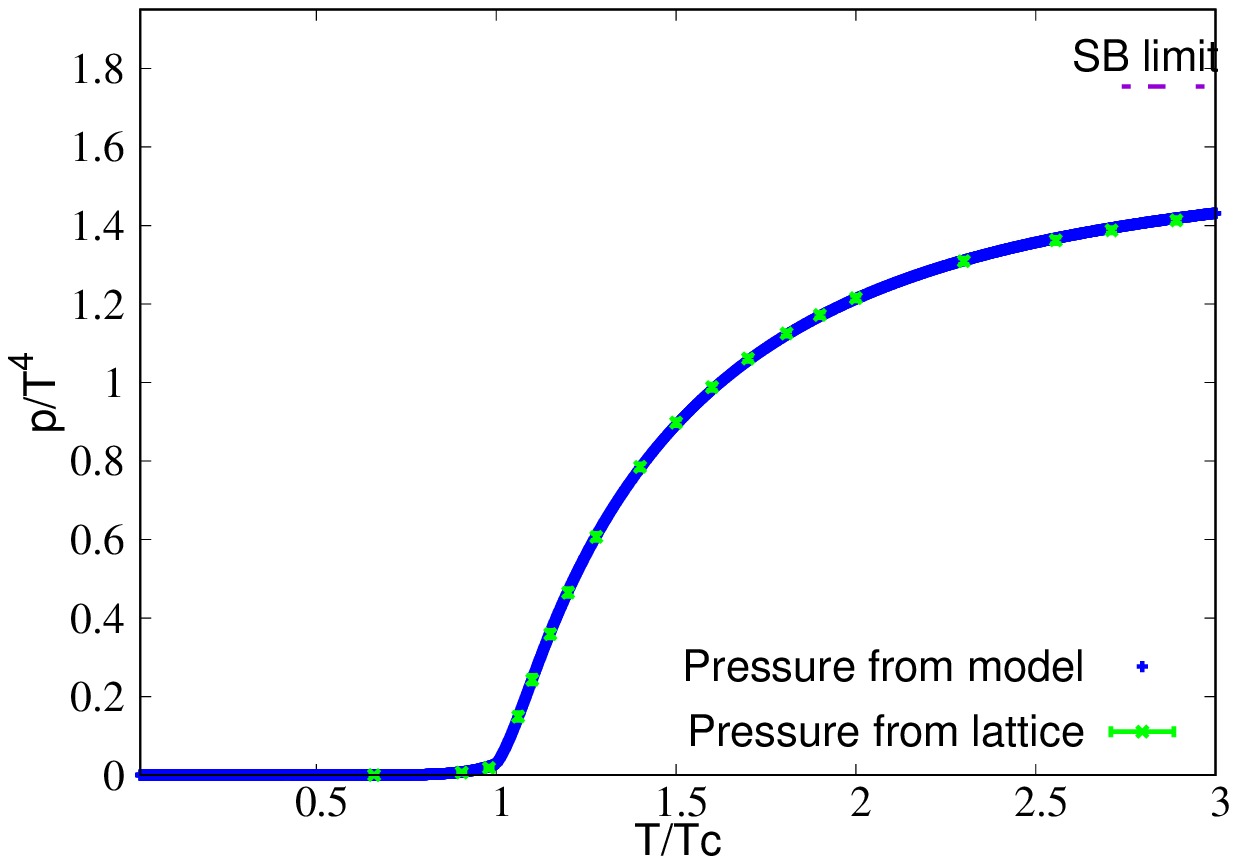}
\caption{Thermal behaviour of pressure without (left) and with (right) mass parameter. Lattice source~\cite{Giusti:2016iqr}.}
\label{fig:pressure}
\end{figure}

\begin{figure}[H]
\centering
  \includegraphics[height=1.5in,width =2.0in]{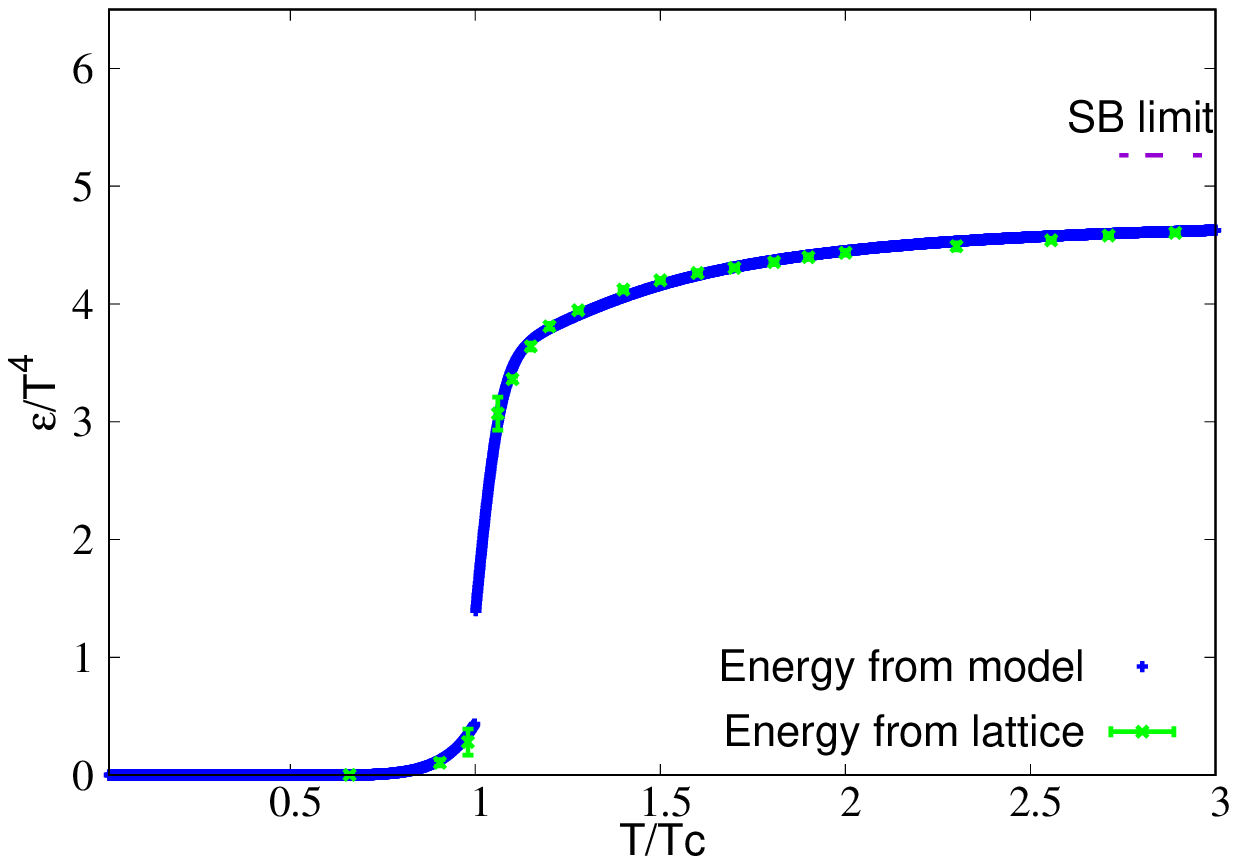}
  \includegraphics[height=1.5in,width =2.0in]{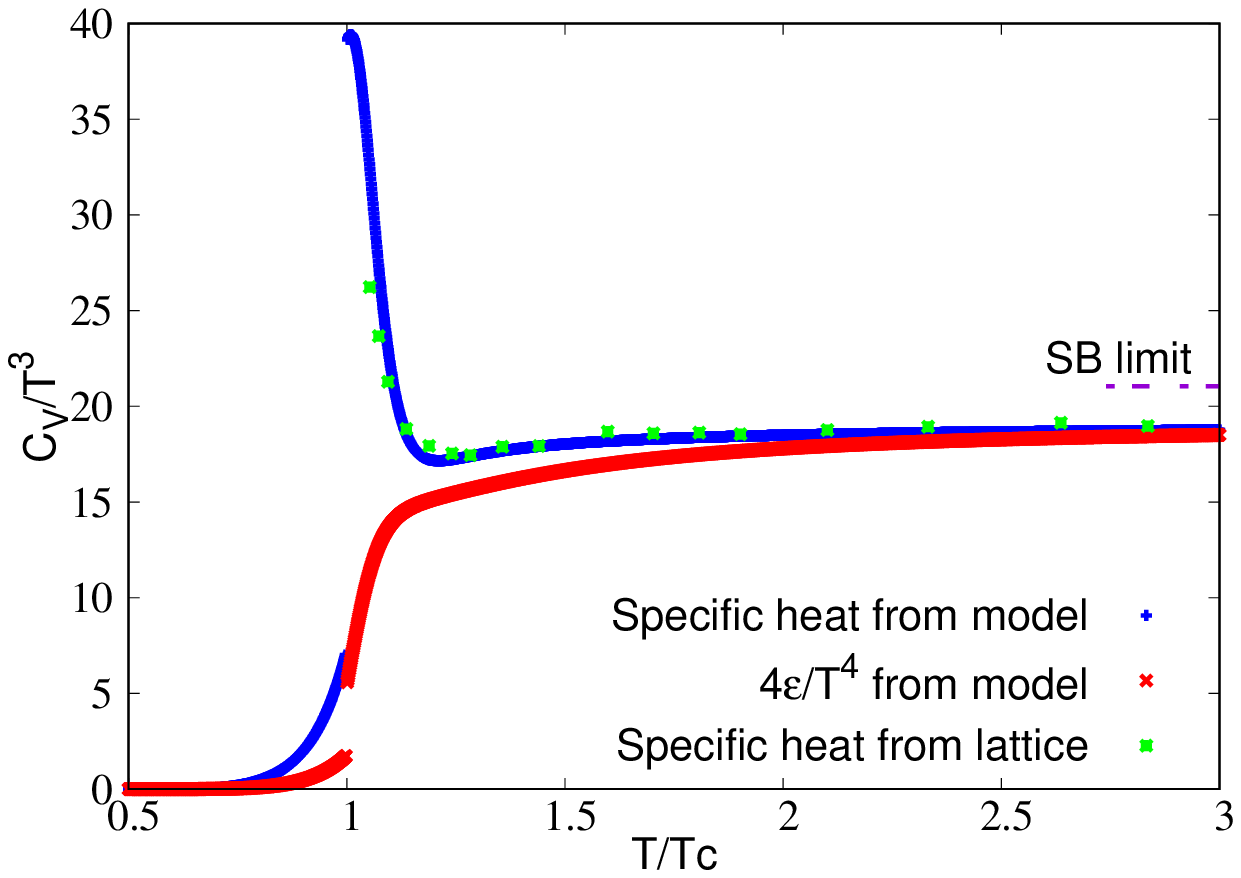}
\caption{Thermal variation of energy (left) and specific heat (right). Lattice source~\cite{Giusti:2016iqr} (left)~\cite{Boyd:1996bx} (right).}
\label{fig:thermo}
\end{figure}

\paragraph*{Parameters:} We choose $\delta v= (0.5T_d)^{-3}$  where $T_d$ is the deconfinement temperature and the obtained pressure shows physically consistent behaviour through out the complete temperature range. However, to have a quantitative agreement to the lattice simulation 
we introduce an effective gluon mass of the form,
\begin{eqnarray}
m_g(T)/T &=& \alpha + \beta/\ln(\gamma~T/T_d), \rm{~for~} T/T_d > 1 \\
         &=& \zeta~(T_d/T)^2, \rm{~for~} T/T_d < 1 ~,
\label{gluon_mass}
\end{eqnarray}
where the parameters are fitted to reproduce the lattice result for the pressure of the SU(3) pure gauge system~\cite{Giusti:2016iqr} and are given by, $\alpha=0.564;\beta=0.176657; \gamma=1.08526\text{ and }\zeta=2.70066$. In the Fig.
(\ref{fig:pressure}), we show the thermal variation of pressure scaled by $T^4$ for both without (left) and with (right) the mass term. Other thermodynamic quantities, derived with the mass term, also show a good agreement with the lattice results (Fig.(\ref{fig:thermo})).

\begin{wrapfigure}{r}{0.5\textwidth}
\centering
    \includegraphics[height=1.3in,width =2.0in]{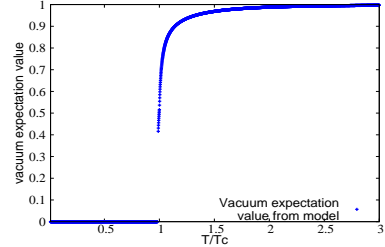}
    \label{fig:saddle}
  \caption{\label{fig:saddle}Thermal behaviour of $\langle\Phi\rangle$. }
\end{wrapfigure}
In the Fig.~(\ref{fig:saddle}), The thermal
behaviour of $\langle\Phi\rangle$, shows a discontinuity at 
transition point, signaling the first order
phase transition.\\  \\ We thus now have a complete
model description for the gluonic medium that can reproduce the symmetry
properties and the thermodynamic observables of an SU(3) pure gauge system. 
\\{\bf Acknowledgment:} This work is supported by DST and DAE of the Government of India. CAI would like to thank TIFR and UCAS for the support.

\end{document}